\documentstyle[preprint,aps]{revtex}

\tightenlines
\begin{document}
\draft
\input{epsf}
\preprint{HEP/123-qed}
\title{Laser induced collisions between Lithium isotopes}
\author{Evgueni V. Kovarski}
\address {ekovars@netscape.net}
\date{07.17.2001}
\maketitle
\begin{abstract}
The near resonant cooling mechanism have offered for
the experiment with laser induced excitation of Lithium isotopes
and its ionization by collisions.
\end{abstract}


\narrowtext
\section {Introduction}

Clearly defined spectrum is the main requirement for any spectroscopical
methods, which deal with two close isotopes as $\it{Li}_6$ and $\it{Li}_7$ ,
Is preferable to use the atomic beams, where it is easy to
realize the excitation of single spectral line  of
$\it{Li}_6$ or $\it{Li}_7$ ~\cite{Sansonetti}. However there are many publications devoted to the
general study of Lithium vapor in the heat-pipe oven. There is the Doppler
broadening of spectral lines.
It is necessary to have a stability and accuracy frequency adjustments of the laser.
Usually for Lithium atom the single mode laser operation is need with stability and tuning
of the laser frequency better then 1 $\it{GHz}$ at the resonant frequency.
The first known publication with Lithium resonance ionization spectroscopy (RIS)
was in 1977~\cite{Karlov}.There were
photo-ionization cross section going from the $\it{2P}$ level and the
demonstration of the ion current by varying the dye laser wavelength
correspondent to $\it{2S-2P}$ transition.
Therefore this was the first work
about the possibility of Lithium laser isotope separation. Two
years later the US patent for Lithium laser isotope separation was accepted in
1979 ~\cite{Yam}. The resonance ionization spectroscopy (RIS) of Lithium
use two lasers with different frequencies. It is depend on intensities of
both lasers for resonance excitation at
$\it{2S - 2P}$ and for the ionization from $\it{2P}$ ~\cite{Arisawa} (1982).
The mixture of two gases in heat pipe usually is present due to chemical activity of
Lithium contacts with optical windows of the heat-pipe oven. The Ar gas usually is present.
Therefore one can to compare ions production by RIS in the heat-pipe with another similar method
for the $\it{Li-Ar}$ contact ionization
based on the ion cyclotron resonance (ICR) isotope separation ~\cite{Suz}.
It should be noted here that we shall not discuss the effectiveness
methods of isotope separation, but on the other hand, both RIS and ICR can work
more successfully at experiments where the concentration of Lithium atoms is high.
To achieve high values of
concentration it is necessary to increase the temperature of $\it{Li}$ of a
vapor chamber with the metallic Lithium.
The ion generation process is important for some applications
at similar condition of environment.
  We consider collisions at the experiment with heat-pipe oven, filled with $\it{Li-Ar}$ mixture
  For such experiment the electromagnetic properties of gas mixture play an
important role due to relatively high concentration of charged particles
produced by  emission, atomic collisions and proper RIS
for ionization. The geometric distribution of generated  positive and negative
particles in the gas chamber is along the way of coaxial laser beams. If there is no
voltage applied to electrodes the small value of electric current is only due to
atom - field interaction. Because the heat pipe oven has two electrodes with
cylindrical symmetry and proper laser is as new cylindrical electrode that
produced charged particles along the laser way.

We are going to present: the simplified method for spectroscopic investigation of the dense
vapor of $\it{Li-Ar}$ mixture, based on methods of atomic collision
estimations; the experimental setup for ionization experiment with lasers,
precisely tuned on transition frequency; the results of electric current
measurements. The ion generation is with both the RIS and without it.
Last process is most interesting to  that the ultra-violet radiation is absent and
for achievement of a threshold of ionization is no sufficiently only to excite $\it{2S-2P}$ transition.
In both cases the quantity of the charged particles can appear so large, that it is possible
to speak about presence low density plasma ~\cite{E1}.
The measurement of the excitation and ionization of dense lithium vapor by cw- dye laser with
wavelength $610.3 \it{nm}$ were performed in Ref.~\cite{Vez}.
In present work  the result was obtained by using only $670.8 \it{nm}$ wavelength.
When the DYE laser excite $\it{2P}$-level at 1,848 $\it{eV}$ it can not reach
to $\it{3D}$ energy level at 3,879 $\it{eV}$  due to 2,031 $\it{eV}$ difference
for two steps excitation and it can lower the ionization potential at 5,391
$\it{eV}$ of $\it{Li}$ atoms. For this
reason we think that there is not a special two-step absorption process, so we will to discuss
two mechanisms such as associative reaction and near resonance cooling.

%%%%%%%%%%%%%%%%%%%%%%%%%%%%%%%%%%%%%%%%%%%%%%%%%%%%%%%%%%%%%%%%%%%%%%%%%%%%%%%
\section{Spectral characteristics of the dense Lithium vapor. }

We begin with note that the main problem for such experiments is
about the relative values for concentrations of
neutral atoms and molecules in the same $\it{Li}$ vapor. Of course, special
measurements for molecules or clusters concentration are needed on time (for
$\it{Li}$ molecules studies see e.g. ref ~\cite{Koch} ).
The standard experimental absorption intensity spectrum of the $\it{2S-2P}$ of each Lithium
isotopes has two spectral lines $\it{D1}$ and $\it{D2}$ ~\cite{Sansonetti} with
correspondence to level isotope shifts~\cite{Niemax}.
  The  Lithium $\it{2S-2P}$ has one doublet with distance of 10
$\it{GHz}$ (0 ,012 $\it{nm}$) between $\it{D1}$ and $\it{D2}$ spectral lines.
The isotope shift between $\it{Li}_{6} \it{D2}$ (670 ,7922 $\it{nm}$) and
 $\it{Li}_{7} \it{D2}$ ( 670 , 7764 $\it{nm}$) is 10 ,5333 $\it{GHz}$ and
between $\it{Li}_{6}\it{D1}$ (670 ,8073 $\it{nm}$) and $\it{Li}_{7}\it{D1}$
(670 ,7915 $\it{nm}$) is 10 ,5329 $\it{GHz}$  ~\cite{Sansonetti}.

Usually the observable at heat-pipe experiment
$\it{2S - 2P}$ spectrum of natural Lithium consists of three lines, where the central
line is formed by the superposition of two lines going from two different
isotopes ~\cite{Yam}. At heat-pipe experiments with the
dense $\it{Li}$ vapor at high temperature, above 1000 $\it{K}$, all Gauss lines
profiles can not been clearly observed. There are only three observable normalized
profiles of absorption intensity received from experiment and fitted
numerically. The well known picture of the fine structure for each isotope was not
at high temperature heat pipe experiment, but particularly was revised with two diode laser for
calibration.

The Voight profile approach can be applied for simple
analysis of the Doppler broadened line and for the Lorenzian line, which widths
are related to the correspondent Gauss to Lorentz $\it{(G/L)}$ damping
constant. With known value of the $\it{(G/L)}$ damping constant from fitted
profile, it is easy to obtain the line width for Lorentzian profile and
evaluate the summary of  excitation transfer mechanisms. Also
special experiments and theory are needed for detailed investigation of
$\it{Li}$ atoms collisions ~\cite{Valda} - ~\cite{Shim}.

We use the method based on dense gas absorption intensity spectrum [Fig.~2] that can be
analytically described by empirical approximation
for a Voight profile of the absorbed intensity in the Beer law,
where the absorption coefficient $\it{k}_{abs}$ is ~\cite{Zemansky}:

\begin{equation}
k_{abs} = k_{G}\left[ e^{\eta}-\frac{ 2 a}{\sqrt{\pi}}
\left(1 - 2\eta F\right)\right] \\ ,
\end{equation}
The absorption coefficient for Gauss line profiles is $\it{k}_{G}$ . The
coefficient $\it{k}_{G}$  can be founded numerically from fitted profiles of
absorption intensity, because the application of the Beer
law with absorption coefficient as a constant, which no depend on Gauss
distribution is no correct for a dense vapor. Also at a maximum of one
unsaturated line is possible to compare $\it{k}_{G}$ form fitted spectrum
with the same value from the Beer law, therefore is possible to use the
thickness (8,5 $\it{cm}$) of dense vapor. There are two numerical
parameters for normalized profile of intensity $\it{I}_{abs}/\it{I}_{0}$, where
$\it{I}_{0}$ is the intensity of the laser. One of these parameters is
$\it{k}_{G}$, which depend on spectral lines saturation, e.g. for $\it{Li}_{6}$
and $\it{Li}_{7}$ due to accordingly concentrations. The $\eta$ is the second
numerical parameter. It dependent on wings of spectral line at frequency $\nu$,
which are far from the central line $\nu_0$ for both isotopes:

\begin{equation}
\eta =
\frac{\left(\nu - \nu_{0}\right)}{\Delta\nu_{G}} \sqrt{l n 2} \\ ,
\end{equation}
For unsaturated spectral lines, as for left spectral profile in the [ Fig.~1 ],
that corresponds to $\it{Li}_6$ , the parameter $\eta$ leads to the function
$\it{F}$ for the Voight profile:

\begin{equation}
 - 2\eta \it{ F} = 1 -
2\eta^{2}+\frac{4}{3}\eta^{2}-\frac{8}{15}\eta^{6}+.. \label{F} \\
\end{equation}
and the $\it{(G/L)}=a$ damping constant is

\begin{equation}
a = \frac{\Delta\nu_{L}}{\Delta\nu_{G}}\sqrt{l n 2} \\ .
\end{equation}
The Eq.~(\ref{F}) is used only for the fitting $\it{Li}_6$ spectral lines,
because the parameter $\eta$ for $\it{Li}_6$ is not too large as for
$\it{Li}_7$, due to the relative concentrations of these isotopes. For
$\it{Li}_7$ isotope spectral line the saturation is much more greater than for
$\it{Li}_6$, thus in this case the Voight profile must be fitted by using the
other sequence:

\begin{equation}
 1- 2\eta\it{ F} =
-\left(\frac{1}{2\eta^{2}}+\frac{3}{4\eta^{2}}+
\frac{15}{8\eta^{6}}+...\right) \\
\end{equation}
By fitting the experimental spectrum with above method
is possible to obtain a better result with experiment for each spectral line. Also
another empirical approximation of Voight profile ~\cite{Whit} can be used
successfully for all spectral lines at the narrow part of frequency spectrum.

 The $\it{(G/L)}$ damping constants for both
$\it{Li}_{6\ D1}$ and $\it{Li}_{7\ D2}$ have the same values
$\it{a}_{7} = \it{a}_{6} = 0.021 \pm 0.0005$ and both absorption coefficients
for $\it{Li}_7$ and $\it{Li}_6$ were estimated from fitted spectrum at a fixed temperature.

The relation of two Gaussian profiles for $\it{Li}_7$ and $\it{Li}_6$ at fixed
temperature is:

\begin{equation}
\frac{\Delta\nu_{G6}}{\Delta\nu_{G7}}=
\frac{\nu_{6D1}}{\nu_{7D2}}\sqrt{\frac{\mu_{7}}{\mu_{6}}}=1,08 ,\\
\end{equation}
where  $\mu_{6}=6,01703\it{a.u.}$, $\mu_{7}=7,01823\it{a.u.}$ and resonance
frequencies corresponds to wavelength 670,7764 nm for $\it{Li}_{7}$ and
670,8073 nm for $\it{Li}_{6}$. At temperature $\it{T}=831.84 \it{K}$ and
pressure $1.8 \it{torr}$, the full width of the half maximum (FWHM) is
$\Delta\nu_{G6}=3763.17 \it{MHz}$ for $\it{Li}_{6\ D1}$ and
$\Delta\nu_{G7}=3484.58 \it{MHz}$ for $\it{Li}_{7\ D2}$.

For correspondent Lorentzian profiles one can easy to calculate the FWHM equal to
$\Delta\nu_{Li6} = 94.92 \it{MHz}$ for $\it{Li}_{6\ D1}$ and $\Delta\nu_{Li7} =
87.89 \it{MHz}$ for $\it{Li}_{7\ D2}$.

The heat velocity of atoms in the $\it{Li-Ar}$ mixture is:
\begin{equation}
\bar {v} =\sqrt{\frac{8k_{B}T}{\pi M}} \\
\end{equation}
Also the numerical equation for corresponded temperature dependence is
 $\bar{v}= 63.64 \sqrt{T}$, where the reduction mass is
 $M = (M_{Li} M_{Ar} /M_{Li} + M_{Ar}) = 0.8683 \cdot10^{-26} \it{kg}$.

Therefore, with the temperature about 650 $\it{K}$ , the random velocity of the
$\it{Li-Ar}$ atoms vapor is about 160 000 $\it{cm/s}$ and the correspondent
Doppler shift for one laser wave is about 16 $\it{GHz}$.
There are some experimental possibilities for measurements.
When the laser has frequency scanning along
spectrum of Lithium, the relative datum  can be used for the
line position in the spectrum due to the Doppler shift for each wavelengths.
The Doppler-free method, that used two opposite
propagated beams with the same frequency on the each laser beam is free of the
Doppler shift for detectors, thus the Lorentzian profile can be experimentally
received, but only calibrated relative scale of the frequency can be used for
estimation of the Lorentzian profile. Therefore the 20 $\it{GHz}$ scale between
known spectral lines captured from the literature datum for $\it{Li}$  is no
good for Doppler-free experimental estimations of the Lorentzian profile in the
gas chamber, because expected value of the FWHM is about 100 $\it{MHz}$ and
special calibration laser air wavelength is needed.
Also if the one of two laser beams
is controlled by wave meter and another laser is controlled by a power meter or
by photo diode with oscilloscope, thus the resolution of the wave meter is about
0,0002 $\it{nm}$ or 133 $\it{MHz}$ at 670 $\it{nm}$ and can not permit well
accuracy for experiments with a Heat-pipe oven where all spectral lines are
Doppler broadened and saturated.
The $\it{2S-2P}$
transition of Lithium isotopes is well known for the all 4 principal spectral
lines  including the fine structures $\it{(fs)}$. Each atom has the lowest
$\it{2S}$ level with two $\it{(fs)}$  levels, which can be observed only at the
special experiment with electric or magnetic fields.
The distance between both $\it{fs}$ levels is 803
$\it{MHz}$ for $\it{2S}$ and 18 $\it{MHz}$ for $\it{2P}$ level.
Yet, we can put fine structures $\it{(fs)}$ out of our consideration the
resonance $\it{2S-2P}$ transition with resonance frequency
$\omega_{0} =2,8\cdot10^ {15} \it{rad./s.}$, because
the Doppler - free
method can not to permit the accuracy experiment with a given transition.

%%%%%%%%%%%%%%%%%%%%%%%%%%%%%%%%%%%%%%%%%%%%%%%%%%%%%%%%%%%%%%%%%%%%%%%%%%%%%%%%
\section {Experimental setup}

  The selection and specification of special designed lasers instrumentation
for selective excitation of lithium isotopes and full ionization is important
because of the small frequency distance between the lithium lines is present
and it is necessary to study some laser  spectral stability
characteristics. The experimental setup [ Fig.~2] consists of three lasers which path
is the same route across the heat-pipe oven. This is possible by use properties of Glan prisms
 and rotators of polarization.

The lasers wavelength was controlled by a serial pulsed Burleigh
wave meter Model WA 4500  with accuracy 0,0001 - 0,0002 nm at wavelength range from 1100 nm to 400 nm .
The laser line shape was measured by use of a
Burleigh Spectrum Analyzer System SA 385  with
SA-91  Etalon Assembly. Its maximum specified resolution is
27 $\it{MHz}$ using 8 $\it{GHz}$ FSR mirror set with finesse about 300  and
the transmission more then 10 %  at 250 - 5000 $\it{nm}$. It can also be used
as a Fabry-Perot filter with thermal sensitivity 70 $\it{nm} / ^{o}\it{C}$
changing in mirrors spacing.

The laser power was measured with a cw - Laser Power Meter,
Spectra Physics, Mod 407 A. It has a wavelength range 250 -1100
$\it{nm}$, power range from 5 $\it{mW}$ to 20 $\it{W}$ with sensitivity
variation about 1 %. It is not designed to monitor $\it{Q}$-switched lasers,
because the damage peak energy density is 0,3 $\it{J/cm}^2$ at 50 $\it{ns}$
pulses. Also the 818 UV, Newport Power Meter was used for precisely measured
power of low intensity cw - lasers.

For the ionization process we used a High Power $\it{Q}$-Switched,
$\it{TEM}_{00}$ frequency-doubled Nd-YAG $\it{Lee}$ Laser, Model
$\it{815 TQ}$. The power stability was 10  % peak-peak. The mean power at the
1064 $\it{nm}$ was 15 $\it{W}$ at 5 $\it{kHz}$ repetition rate
(pulse energy 2 ,2 $\it{mJ}$ and pulse duration 90 $\it{ns}$) and the intensity
3 $\it{MW/cm}^2$. In this configuration the second harmonic wavelength (532
$\it{nm}$) is emitted laterally from the optical axis of the laser resonator.
The SHG assembly contains a KTP (potassium titanyl phosphate crystal) which is
bounced inside a temperature controlled chamber. Because of the small crystal
cross section (4 x 4 $\it{mm)}$ ), there is a little box for misalignment of the
crystal with the Nd-YAG beam optical center line. The average power was 2,5
$\it{W}$, the average energy was 0,5 $\it{mJ}$ and the pulse duration was 100
$\it{ns}$. We worked with 1,5 $\it{W }$ for
obtaining UV 266 $\it{nm}$. For this purpose we used an $\it{INRAD}$
temperature stabilized crystal of
$\it{KDP}$ , obtaining 40  $\it{mW}$ of the average power. The non-critical
phase matching temperature (35 $^{o}\it{C}$ - 75 $^{o}\it{C}$) is quite
dependent on the deuteration level of the $\it{KDP}$. Also the incoming edge of
the crystal is at 90 $^ o$  with the optical axis of the crystal and the
reflection at the angle about 23 $^ o$ decrease the intensity of 532 $\it{nm}$
about two times. The lineal polarization of 532 $\it{nm}$ going from the
$\it{LEE}$ laser had an inclination with the angle 45 $^ o$  respect to the
optical path of our experiment. The specially oriented crystal $\it{KDP}$ for
second harmonic generation $(\it{SHG})$ was needed. From two waves within one
laser beam with $\lambda$ = 532 $\it{nm}$  the well known procedure of the
second harmonic generation was used. The wave with $\lambda$ = 266 $\it{nm}$
We obtained a UV-beam at 266 $\it{nm}$ with intensity 16 $\it{kW/cm} ^2$ after the prism
separation from incoming 532 $\it{nm}$ wave. The line width of
the UV line was 20 $\it{MHz}$ with high frequency stability.
Frequency stability of the ionization laser is no so critical as for the excitation process, because  for
ionization we have a transition from $\it{2P}$ to above ionization limit.

The group of calibration lasers consist both the 6202 $\it{New Focus}$ diode
laser and the $\it{EOSI}$ diode laser with
line width about 100 $\it{KHz}$ and the average power 6 $\it{mW}$. These lasers
was used for calibrating the experiment due to high stability parameters at
the needed wave length.

We tested a cw - broadband
standing-wave Spectra Physics DYE Laser (Model 375 B), because we can change
the mode structure of outgoing  laser wave as a special background for given laser
equipment ~\cite{E5}. The DYE laser was pumped by a power $\it{P}_{0}=6 \it{W}$ ,
Stability Multiline Ar-Ion Laser Spectra Physics Model 2017 -06 S. The Ar-ion laser had some measurable
spectral lines, such as: at 514,5 $\it{nm}$ with 0,393 $\it{P}_{0}$,
at 496,5$\it{nm}$ with 0,166 $\it{P}_{0}$, at 488 $\it{nm}$ with 0,285
$\it{P}_{0}$, at 476,5$\it{nm}$ with 0,109 $\it{P}_{0}$, at 457,9 $\it{nm}$ with
0,047$\it{P}_{0}$. The amplitude stability of the DYE laser intensity
was best with pumping of the Ar-Ion laser at 514,5 nm. The DYE laser was based on DCM
4 -$\it{(Dicyanomethylene)}$ - 2 -$\it{(methyl)}$ - 6
-$\it{(p-dimethylaminostyryl)}$- 4-$\it{H}$-$\it{(pyran)}$ with molecular
weight 303,37 $\it{a.u.}$ and concentration 1,5 $\it{(mmole/l)}$ , which was
prepared when 0 ,682$\it{g}$ of the DCM was solved in Prophylene carbonate (0
,6 $\it{l}$) and Ethylene Glycol (0,9 $\it{l}$).
The absorption curve of DCM has
an absorbency region between 450 $\it{nm}$ and 550 $\it{nm}$ . The broad band
emission curve has the FWHM about 80 $\it{nm}$ at 650 $\it{nm}$,
which is close to the resonance $\it{2S-2P}$ transition of Lithium at 671
$\it{nm}$. The Spectra Physics data for the Model 375 B laser were: power 600 $\it{mW}$ at
705 $\it{nm}$, line width 7 $\it{GHz}$, amplitude stability less then 0 ,5 % ,
accuracy of wavelength changing 0 ,5 $\it{nm}$.
The limit of wavelength changing is in the same order as a
Rayleigh selection rule for spectral lines $ 8/\pi^{2}$, that is about 0,5
$\it{nm}$ for a given laser line width.
We obtained characteristics of
Spectra Physics DYE laser, Model 375 B, such as: average power 1 ,2 $\it{W}$
at 671 $\it{nm}$ , accuracy of changing wavelength better then 0,01 $\it{nm}$
and frequency stability about 0,0001 $\it{nm/hour}$ during time of
the experiment.  All parameters are depended on needed laser modes structure
and corresponds to properties of optical elements inside the resonator.
For our experiments with $\it{Li}$ we modified the mode structure and the spectral line width of
outgoing laser wave by inserting an additional etalon into a laser resonator.
We note that
broadband laser had the tuning accuracy 0,5 $\it{nm}$ without $\it{E1}$ due
to $\it{BF}$ frequency tuning accuracy. With $\it{BF}$ and $\it{E1}$ the
frequency tuning accuracy was measured $\approx$0,0001 $\it{nm}$. The first
air spaced etalon $\it{(E1)}$ had a temperature stability. When the $\it{BF}$
and $\it{E1}$ were used, the output frequency of the DYE laser was tuned to the
quite frequency 4,46789597$\cdot 10^{14}\it{Hz}$ by used the angle and
temperature tuning of $\it{E1}$. The FSR of output frequency was 19,655
$\it{GHz}$. Then the second etalon $\it{E2}$, with FSR = 15 $\it{GHz}$ was
installed inside the cavity and the same operation was repeated with angle
tuning of $\it{E2}$ for the same frequency. The $\it{E2}$ was chosen, because
the spectral lines of the Lithium have 10 $\it{Ghz}$ distance twice between 3
spectral lines and laser with $\it{BF}$ and $\it{E1}$ can excite both
$\it{Li}_{6}$ and $\it{Li}_{7}$. The single mode operation of the broad band
laser was achieved.  The single mode selection was realized experimentally [Fig.~2]
 by use the aperture with diameter 1 $\it{mm}$
for transverse mode selection and, by use two etalons $\it{(E1,E2)}$ and the
birefringent filter $\it{(BF)}$ for longitudinal mode selection.
The mode structure of the single laser beam is very important for selective
excitation. There are some experimental possibilities with the broad band DYE
laser, which depend on needed experiments. The sngle mode operation was necessary for
a RIS experiments. On the contrary for measured values of electric current
[Fig.3] of the laser induced low density plasma observation
experiment ~\cite{E1}, the DYE laser had the two longitudinal
modes with FWHM 100 $\it{MHz}$, separated exactly at
20 $\it{GHz}$ with cavity etalon. This etalon had the free spectral range (FSR)
that correspondent the distance between two spectral lines of $\it{Li}_{6}$ and
$\it{Li}_{7}$ isotopes of the natural Lithium vapor. The laser was tuned
exactly on 670.8073 $\it{nm}$ for $\it{Li}_6$. Another wavelength 670.7716
$\it{nm}$ was not controlled. The average power at this experiment was 100
$\it{mW}$. There are some known methods of longitudinal mode selection by use interferometers
techniques as in ~\cite{Smith},~\cite{Skowronek}.
The free spectral range FSR is equal to the distance between two longitudinal
modes:

\begin{equation}
 F S R_{\nu} = \frac{c}{2\eta_{t} d \cos\Theta_{i}}  \\
\end{equation}
Thus, with variation of the angle, some longitudinal modes of the laser will have
not losses for correspondent frequencies $\nu = \it{m}$, (FSR), and it is
possible to move the Airy spectrum of the etalon along the frequency axis up or
down to quite position. The longitudinal modes are
spaced in frequency by $\it{c/2L}$. We note that for near-confocal resonator
the free spectral range (FSR) is:

\begin{equation}
\delta\nu = \frac{c}{4 L + \frac{x}{\rho^{3}}},  \\
\end{equation}
where $\it{(x)}$ - is the distance from the axis and $\rho$ is the radius of
mirrors, when $\it{L} = \rho $. Thus the series of longitudinal modes are
separated by $\it{c/4L}\approx 250 \it{MHz}$ for a cavity with
$\it{L}\approx 30 \it{cm}$. Transverse modes are important for a mode selection too.
Amplitude distribution of these modes is given with Hermit polynomial
$\it{TEM}_{ m n}$ modes. The separation of the first transverse mode ($\it{m = 0, n = 1}$) from the
 axial mode is:

\begin{equation}
\frac{\nu_{1}-\nu_{0}}{\nu_{0}}  = 2,405\frac{\lambda^{2}}{ 8\pi^{2} r^{2}},   \\
\end{equation}
where $\it{(r)}$ is  diameter of cylindrical laser material.
  In deriving the Van Cittert-Zernike
theorem ~\cite{Bor} it follows that the diffraction pattern would be obtained
on replacing the source (the DYE luminescence cross section) by a diffraction
aperture of the same size and shape as the source. The amplitude distribution over the wave front
in the aperture being proportional to the intensity distribution across the source.
The calculation of the complex degree of coherence of light from an incoherent
source, by use the Hopkins formula ~\cite{Bor}, leads to the mirrors diameter
0,32 $\lambda\it{f / h}$, where $\it{(f)}$- is focal distances ,
$\it{h}\approx 10 \mu \it{m}$ - is the diameter of the source, that dependent
on a pumping of the focused laser (6 $\it{W}$ Ar-Ion laser) and thermal lens. At
flow velocities of the DYE about 10 $\it{m/s}$ the time of flight for the dye
molecules through the focus is about 1 $\mu\it{s}$. By using a mirror whose
reflectively can be varied over its surface, it is possible to have some given
combination of transverse modes, if it is necessary.

There are two problems of frequency stability of the DYE laser.
Firstly, we note that the material of installed cavity DYE laser etalon
was fused silica. Regardless of thickness it has a drift of airspace between
fused silica plates about 5 $\it{GHz}/ ^{o}\it{C}$ and would have to be
temperature stable about 0,1 $^{o}\it{C/hour}$.
Secondly, the environment conditions are very important for stable laser
experiment with atomic spectral lines. Therefore, there is the frequency drift:

\begin{equation}
\frac{\Delta\nu}{\nu} = \frac{\Delta d}{ d} + \frac{\Delta n}{ n} \\
\end{equation}
where ($\it{d}$) is the cavity length and ($\it{n}$) is the refraction
coefficient.
For the quartz rod resonator structure of the DYE laser,$\it{Model 375 D}$, we
have with respect to the change of a cavity length a resulting temperature
sensitivity about 90 $\it{MHz}/ ^{0}\it{C}$. With  respect to the refraction
coefficient of air we found a temperature sensitivity about
410 $\it{MHz}/^{o}\it{C}$. The amplitude and frequency stability of the laser was maintained
during 3 hours of the experiment. The wavelength control had the accuracy
0.0001 $\it{nm}$ by use the serial wave meter. The measured value of the line width
was about 100 $\it{MHz}$ for
cw -regime of DYE excitation by use the focused $\it{Ar-Ion}$ laser.
Another excitation of the same DYE laser was with 120 $\it{ns}$ pulses of twice doubled
  Nd:YAG laser at 266 nm, with 5 $\it{KHz}$ pulse repetition and average power about
  1 $\it{W}$. Output pulses from the DYE laser were 100 $\it{ns}$.

%%%%%%%%%%%%%%%%%%%%%%%%%%%%%%%%%%%%%%%%%%%%%%%%%%%%%%%%%%%%%%%%%%%%%%%%%%%%%%%%
\section{Experimental result}

For a known value of the $\it{(G/L)}$ damping constant and absorption
coefficient from described method of the fitting we can exam the concentration
of the both isotopes:

\begin{equation}
 N_{6G} =
k_{6G}\left[\frac{2\sqrt{\pi}}{\lambda^{2}_{6D1}}\right]\frac{1}{a} \\ ,
\end{equation}
\begin{equation}
N_{7G}= k_{7G}\left[\frac{\sqrt{\pi}}{\lambda^{2}_{7D2}}\right]\frac{1}{a} \\ ,
\end{equation}
where the absorption coefficients given from fitted profiles are
$\it{k}=0.6\it{cm}^{-1}$ for $\it{Li}_6$ and $\it{k}=14.75 \it{cm}^{-1}$ for
$\it{Li}_7$. The correspondent concentration going from above formulas is equal
to $2.25\cdot10^{10}\it{cm}^{-3}$ for $\it{Li}_6$ and
$27.65\cdot10^{10}\it{cm}^{-3}$ for $\it{Li}_7$. So the relation of two
concentrations is $\it{N_6/N_7}$ = 0.08135, which is in well agreement with
the characteristics of natural Lithium ~\cite{Shin}.

On the other hand, one can to verify values of Lithium concentration by use the
following well known formula for the Lithium vapor
pressure (in $\it{torr}$) at fixed temperature (in $\it{K}$):

\begin{equation}
 P = 1 0^{5,055 -\frac{8023}{T}} , \\
\end{equation}
where the vapor pressure for given temperature is equal to
0.003427 $\it{Pa}$ = 0.0000257 $\it{torr}$. By using the equation of the state
for ideal gas $\it{P=N k}_{B}\it{T}$ and Eq.(10 ), one can obtain the
concentration of the natural Lithium, which is equal to
$29.85\cdot10^{10}\it{cm}^{-3}$. The Heat-pipe was filled with $\it{Ar}$ gas
with the concentration $10^{17}\it{cm}^{-3}$.

When resonance excitation of the considered system
$ \it{Li} (2^ {2} S)  -(2^ {2} P)$  was realized by 671 $\it{nm}$ cw-laser, the
ionization process was registered. The result of ionization by 671 $\it{nm}$
was compared with additional UV laser for ionization from $\it{2S - 2P}$ energy
level (1,848 $\it{eV}$) to energy level of ionization limit (5,391 $\it{eV}$).

Intensity values of both the cw-DYE laser and the pulsed UV laser
were changed correspondingly
to the beam diameter 3 $\it{mm}$ across the heat-pipe oven. We
did not worked with specially applied to the heat-pipe electrodes voltage or
load resistors.

 Direct measurements of the electric current were performed by
amper meter; maximum average value of the electric current was registered
about $30\mu \it{A}$ [ Fig.~3 ]. We denoted three curves, where three principal effects were
registered such as:
(a) is the upper dependence for the RIS by use excitation and ionization from
two lasers $671 nm$ and $266 nm$. The RIS is more effective between $450^o C$
and $650^o C$.
(b) is the ionization current due to only $2S$-$2P$ excitation with $671 nm$.
This specific ionization is more effective for highest temperatures.
(c) is the lowest dependence (the noise) for a  emission.
Only small difference between average
value of electric current due to the RIS process [ Fig.~3 -(a)] and to resonance excitation
$\it{2S-2P}$ [ Fig.~3 -(b) ].
The noise or the
emission in temperature range from $450^o \it{C}$ up to $700^o \it{C}$ with
electric current up to $3 \mu \it{A}$ and capacitance up to $20 \it{nF}$  with
observable saturation [ Fig.~3 -(c) ].
Values of the electric current varies from $5 \mu A$ up to $30 \mu A$ [ Fig.~3 -(a)],
and capacitance from $10 \it{nF}$ up to $100 \it{nF}$ .
Also the capacitance of cylindrical heat-pipe
 was measured. The results of the
current and capacitance measurements at the
temperature higher than $650^o \it{C}$ are the same for both mechanisms.
Also it is easy to
separate RIS from another ionization using the modulation signal measurements because
RIS use the pulse UV laser.

%%%%%%%%%%%%%%%%%%%%%%%%%%%%%%%%%%%%%%%%%%%%%%%%%%%%%%%%%%%%%%%%%%%%%%%%%%%%%%%%

\section {Some mechanisms of collisions}

  If the colliding particles have no charge, the trajectory $\it{r(t)}$ is
assumed to be rectilinear: $\it{r(t)} = \rho + \it{vt}$ ,  where $\rho$ and
$\it{v}$ are the impact parameter and relative velocity. The elastic collisions
change the phase of damped oscillator, $\phi (t)$, but not the amplitude,
$\it{x(t)}$, due to the frequency shift $\Delta\omega$ during the
phase-perturbing collisions. If the duration of collision, $\tau_{c}$, is short
compared with the mean time between two consecutive collisions, one can neglect
the radiation during collisions and consider the collisions to be
instantaneous. Therefore the collisions are manifested only in phase shifts
(Impact broadening approximation), where the dimension parameter
 $h = \rho^{3}\it{N}_{Ar}\ll 1$ determines the number of perturbers in the
Weisskopf sphere with the radius $\rho$, where the binary approximation holds.
Depending on the type and states of interacting atoms both attraction and
repulsion can take place at long distances. At short distances the potential is
repulsive. In general case the interaction force is depends not only on spatial
but also on angular variables. The results of calculations in which a more
realistic interaction $\it{r}^{-6}$ is used can not be described by a simple
Lorentzian distribution, that depends on the type of the transition. The
repulsive part of the interaction is usually taken into account in the form of
Lennard-Jones potential. In general case, when the mean distance between the
atoms is of the same order with the magnitude of atomic dimensions, simple
expression for potential $\it{V(r)}=\it{C}_{6}/\it{r}^{6}$ is valid with
approximations due to both instantaneous dipole-dipole interaction and
dipole-quadruple interaction with $ \it{C}_{8}/\it{r}^{8}$.

The shift, $\Delta\omega = 2\pi\Delta\nu$, and broadening, $\Gamma$, of the Lorentz
spectral distribution~\cite{Sobelman} are:

\begin{equation}
\Delta\nu = 2.91 C^{\frac{2}{5}}_{6} v^{\frac{3}{5}} N_{Ar}  \ ,
\end{equation}

\begin{equation}
\Gamma = 2 \pi\Delta\nu_{L} = 8.16 C^{\frac{2}{5}}_{6} v^{\frac{3}{5}} N_{Ar}
\ ,
\end{equation}
For a temperature $T=831K$, which we used for a fitting procedure, one can
easily obtain the value for a FWHM of the Lorentzian profile,
$\Gamma = 617 \it{rad} \cdot \it{s}^{-1}$ or $\Delta\nu = 98.2 \it{MHz}$
by using the coefficient in the length form
$\it{ C}_{6}$=1406 a.u.~\cite{Marinescu} for $ \it{Li} (2^ {2} S)  -(2^ {2} P)$
, where two interacted atoms, one of which $2^2\it{P}$,  are in different
angular momentum states with associated magnetic quantum numbers
$ \it{M}_{2}= \pm 1$ and attractive or repulsive interaction are.
Also for $\it{Li} (2^ {2} S) - \it{Li} (2^{2} S)$ the  value of
 $ C_{6}$=1390 a.u. = 1390
 $ e^{2}a^{2}_{0} = 1.330786 \cdot10^{-76} Cl \cdot V \cdot m^{6}$ ~\cite{Fish}
 is closely to estimation of $\it{ C}_{6}$ for
$ \it{Li} (2^ {2} S)  -(2^ {2}  P)$   interaction. For this reason the measured
ionization gives the answer about the nature of interaction in favor of
$ \it{Li} (2^ {2} S)  -(2^ {2}  P)$ .

Therefore the value of $\Delta\nu = 98.2 \it{MHz}$  which can be obtained by
use$\it{C}_{6}$ coefficient well corresponded to experimental results
$\Delta\nu_{L6} = 94.92 \it{MHz}$ for $Li_{6\ D1}$ and $\Delta\nu_{L7} = 87.89
\it{MHz}$ for $\it{Li}_{7\ D2}$
from the fitting procedure.

The shift of the spectral maximum position measured at two
temperature points are $0.0126 \it{MHz/K}$ or $1.26 \it{MHz/torr}$.
The value of the FWHM temperature deviation is
$0.0353\it{MHz/K}$ due to the principal temperature dependence
 $\Delta\nu\simeq \it{T}^{0,3}$, which implies the correspondent pressure
 changing as $3.53 \it{MHz/torr}$ (for a comparison see results for $\it{Li}$
in ref.~\cite{Demtroder}).

The impact parameter or Weisskopf radius is~\cite{Sobelman}:

\begin{equation}
\rho = \left(\frac{3\pi}{8}\frac{C_{6}}{v}\right)^{0,2}
\end{equation}
For a temperature range from $500^o \it{C}$ to $750^o \it{C}$
the impact parameter varies from $6.685\AA=12.63a_{0}$ to $6.5\AA =12.28 a_{0}$.
The cross section is
$\sigma\simeq\pi\rho^{2}=132.7\AA^{2}=1.33\cdot10^{-14}\it{cm}^{2}$.

The minimum concentration of
charged particles in $\it{Li-Ar}$ mixture at temperature 1000 $\it{K}$ can be
estimated by taking into account the value of total current between two
electrodes of Heat pipe as $\it{n}_e \approx 10^6\it{cm}^{-3}$. Well-known Saha
equation for amount of ionization $\it{n_e/n}$ to be expected in a gas in
thermal equilibrium

\begin{equation}
\frac{n_e}{n} \approx 2.4 \times 10^{15} \frac{T^{3/2}}{n_e} e^{-U_i/kT}
                                                   \label{Saha}
\end{equation}
gives the ionization degree $\approx 10^{-10}$ for $\it{Li}$ (which
concentration is $\it{n}_{Li}\approx10^{13}\it{cm}^{-3}$ and ionization energy
$\it{U}_i=5.4 \it{eV}$) and $10^{-36}$ for $\it{Ar}$ ,
($\it{n}_{Ar}\approx10^{17}\it{cm}^{-3}$, $\it{U}_i=15\it{eV}$). So, the
inelastic collisions between unexcited $\it{Ar}$ and $\it{Li}$ atoms could not
provide the observed concentration of charge particles.

Such value of electron gas density, when DYE laser is applied [ Fig.~3 -(b) ], can be
achieved due to the yield of the associative ionization reaction:

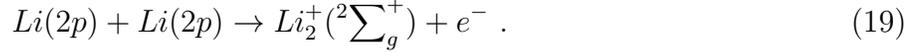
\begin{equation}
Li(2p) + Li(2p) \rightarrow Li_2^+(^2{\sum}^+_g) + e^- \ .
\end{equation}
Such a process would be possible if the excited lithium atoms approached each
other with relative kinetic energy about of $0.46 \it{eV}$  \cite{Kon}.

The simple approximation (\ref{Saha})
shows that above associative ionization reaction can provide the electron
densities up to $\it{n}_e \approx10^8\it{cm}^{-3}$ without the action of UV
laser [ Fig.~3 -(b) ]. As for the rather high current observed even in the case when both lasers (DYE
and UV) are shut down [Fig.~3 - (c)], such value of electron gas density
can be achieved due to the effects of  emission, mainly from the
internal hot electrode with a mesh plating by Lithium.

  The space-charge average current in Heat-pipe can be estimated by using the
Child equation~\cite{Smyth}, due to the presence of cylindrical symmetry:

\begin{equation}
- I = \frac{8\pi\epsilon}{9}\sqrt{\frac{2e}{m_e}}\frac{V^{3/2}}{a\beta^2} \ ,
\end{equation}
where the parameter ($\beta$) was determined in the Ref.~\cite{Smyth} and
($\it{a}$) is the radius of the external cylinder. The mean values for current
and dielectric constant can be taken directly from current and capacitance
measurements.

Given the natural line width equal to 5,8 $\it{MHz}$ for the $\it{2P}$ - energy
level we obtain the absorption cross section

\begin{equation}
\sigma_{ 2 s - 2 p} = 0,054   A _ { 2 p - 2 s}   \frac{\lambda
^2}{\Delta\nu_{ D}} \\
\end{equation}

where $ \it{A}_{ 2 p - 2 s}$ = 1 /$\it{t}_{s}$ , with $\it{t}_{s}$  = 27 ,1 ns. The
Doppler line width of the $\it{Li}$  was $\Delta\nu_{ D} $ = 2,5 GHz. The
absorption coefficient was 0,154 $\it{cm }^{-1}$  in our Heat-Pipe
experiment. For a given $\it{2S-2P }$  resonance excitation we measured
absorption cross section

\begin{equation}
\sigma _{ a} = 1 , 4   . 1 0  ^{ - 1 2} c m ^ 2 \\
\end{equation}
and ionization cross section for 266 nm ionizer:

\begin{equation}
\sigma _{ i} = 1 , 2   .  1 0  ^{ - 1 7} c m ^ 2   \\
\end{equation}

The cathode emission current and the current of
positively charged ions were also discussed in the Ref.~\cite{Thomp}.

The one-dimensional systems of driven charges can change symmetry in the phase pulses space,
when the classical phenomena in one direction take into account
quantization in others. It results in thermal drag effects in solids wires ~\cite{E16}, but can result
in thermal optical experiments too. Another effect is light-induced diffusive pulling (pushing)
of lithium atoms into a laser beam, which was studied in Ar noble gas, where the value of
diffusion coefficient 4.3$\it{cm}^2/\it{s}$ was measured ~\cite{At}.

It should be noted here, that the peak concentration of charged particles
produced by the RIS [ Fig.~3 -(a) ] using 77 $\it{ns}$  UV pulses is much more higher than the
measured average value.

An estimation, given even for the minimum value of
$\it{n}_e \approx 10^6 \it{cm}^{-3}$, gives the value of Debye length
 $\lambda_D \sim 0.1 \it{cm}$, the number of charged particles inside the Debye
  sphere $\it{N}_d \sim 10^5$ and $\omega_p \nu_{ei} \sim 10^4$,
where $\omega_{p}$ is the plasma frequency and $\nu_{ei}$ is the
collisions rate. It means that the $\it{Li-Ar}$ mixture at temperature in the
order of $10^3 \it{K}$ behaves like a plasma rather than a neutral gas  ~\cite{E1}.

By other hand we can discuss another mechanism for observed ionization .
\\
We will take the Doppler cooling as the one step to achieve the low temperature
for the Bose-Einstein Condensate (BEC) at very low temperature.
The limit of the temperature achieved for Lithium is 140 $\mu\it{K}$
by balance with a scattering force in a random direction~\cite{Hulet}.
The Doppler limit of achievable temperature $\it{(T)}$ is:

\begin{equation}
 k_{B}\cdot\it{T} =\frac{\hbar\Gamma}{ 2}
\end{equation}
where $\Gamma$ is a damping constant.

Let's change this relation:

\begin{equation}
k_{B} T =\hbar\cdot\frac{\it{R}}{R}_{0}=\hbar\cdot\it{R}_{1}
\end{equation}
where the rate of the power absorption $\it{R}_{1}$
for two level atomic system interacted with EM field is well known:

\begin{equation}
 R_{1} =\frac{ 2 \Omega^{2}\Gamma}{\Delta\omega^{2}+
 4\Omega^{2}+\Gamma^{2}} \\
 \end{equation}
The Rabi frequency $\Omega_{\nu}$ measured in ($\it{MHz}$) is:

\begin{equation}
\Omega_{\nu} = \frac{\it{d}_{1,2}\cdot\it{E}_{0}}{\hbar}  \\
\end{equation}
with $\it{d}_{1,2}$ - as the matrix dipole momentum and $\it{E}$ as the amplitude of EM wave.
There is the frequency $\Omega$, that make a round oscillations between two levels.
When $\Delta\omega\not=0$ , one can used
for a given $\it{Li}$ atom the fixed value of the Rabi frequency 1000
$\it{MHz.rad.}$. Some inequalities are for given values in the denominator of
the profile. One of this inequalities is evidently for an application,
because the value of the Rabi frequency $\Omega=\Omega_{c}$ =1000
$\it{MHz.rad.}$ and $\Gamma=  37 \it{MHz.rad.}$ with $\Omega\gg\Gamma$ :

\begin{equation}
 R_{1}=\pi\Omega_{c}\Gamma\left[\frac{1}{ 2\pi}
 \frac{ 4\Omega_{c}}{\left(\omega-\omega_{0}\right)^{2}+
 \left( 2\Omega_{c}\right)^{2}}\right] \\
\end{equation}
The rate $ R_{1}$  can be represented by
use the parameter $\rho$   that depend on the Rabi frequency and the
frequency tuning value:

\begin{equation}
\rho = \frac{\Omega}{\Delta\omega} = c o n s t \left[\frac{\sqrt{
I}}{\Delta\omega}\right] \\
\end{equation}
\begin{equation}
 R_{1} =\frac{\Gamma}{2}\left[\frac{( 2\rho)^{2}}{1 + ( 2\rho)^{2}}\right] \\
\end{equation}
 The well known saturation effect by intensity of the laser is clearly observed.
Is possible to fix at experiment both the frequency $\Delta\omega$ and the laser
intensity $\it I$. Therefore, the correspondent $\rho$ must depend on the intensity [ Fig.~4 ]
or on the frequency [ Fig.~5]. The simplest calculation show that the saturation is at
 $\Delta\omega\simeq$ 0,02 $\Omega$ and absorption begin at $\Delta\omega\simeq\Omega$.
As the $\Gamma$ is return size of the time,
it is possible to transfer a condition of saturation to language of the time-table interaction
~\cite{E9}. At the fixed frequency [ Fig.~4 ], when increase of intensity of the laser, the saturation
comes at the maximal size of $\Gamma$, that corresponds to the minimal time.
Very important the fact of weak growth is that means that the time too gradually
decreases aspiring to a limit. On the other hand at a weak intensity the time is very large
also grows indefinitely.
At the fixed size of intensity of the laser [ Fig.~5], the condition of an exact resonance coincides
with a condition when time is equally practically to zero and then it is gradually increased
to indefinitely in process of a distance from a resonance ~\cite{E9}.
If now to put in conformity a parity between frequency $\Delta\omega$ of a field and $\it{k}\cdot\it{v}$
where is the speed of atoms, becomes understandable that now it is possible to speak not only
about time of interaction or time of life ~\cite{E9},
but also about speeds of atoms, because it is supposed that is possible to replace one ($\Delta\omega$)
with another ($\it{k}\cdot\it{v})$ accordingly. However, if we can put ($\Delta\omega$) as a constant,
another $\it{v}$ is only one part of the distributions function.

When the condition for the near resonant condition is
$\Delta\omega\gg\gamma$ for the quantum interference condition of the EM field,
then the rate $ \it{R}_{2}$ is simply ~\cite{E8}:

\begin{equation}
 R_{2} = \frac{\Gamma}{2}\left[1 - J_{0}\left( 2\rho\right)\right],  \\
\end{equation}
where $\it{J}_{0}$ is a Bessel function. Practically situation is as
for one frequency, but with important features [ Fig.~6] due to a Bessel function.

With relation between $\it{k}\cdot\it{T}$ and the rate $\it{R}$ all figures becomes
clearly observed due to connection of the temperature with the atomic velocity.
Other atoms have conserved high velocities. Therefore the inelastic collisions should be expected.
 The general velocity distribution can be changed and the satellite distribution
can be observed due to laser induced collisions between atoms.

\section{Figure Caption}

Fig.1.
Lithium 2S-2P absorption normalized spectrum with fitting

Fig.2.
Experimental setup for RIS of Lithium isotopes with Heat-Pipe oven and
lasers.

Fig.3.
The electric current measurement.

Fig.4
Saturation of the rate at the fixed frequency.

Fig.5
The rate at the fixed intensity.

Fig.6
The quantum interference rate ar the fixed frequencies difference.

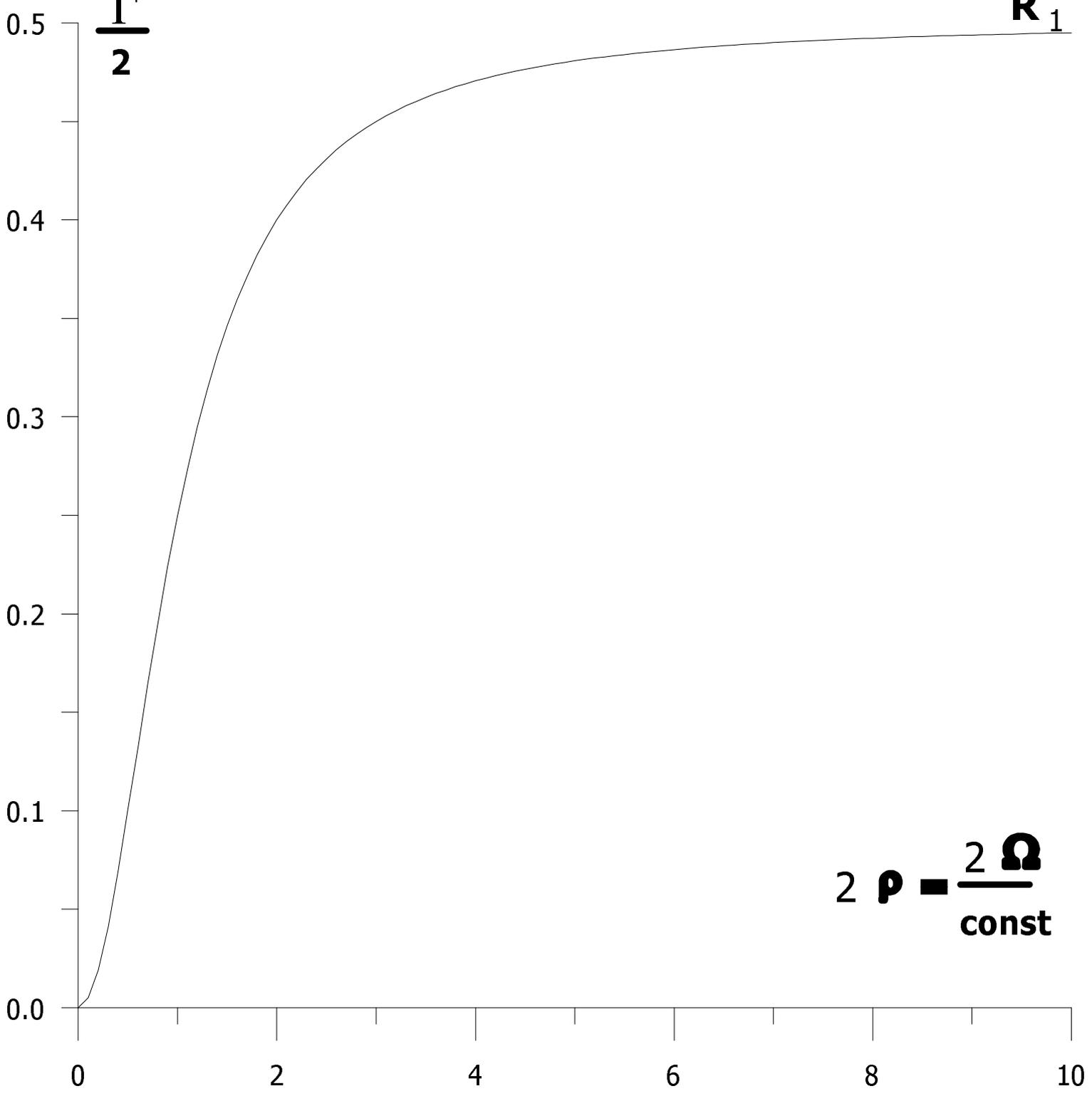
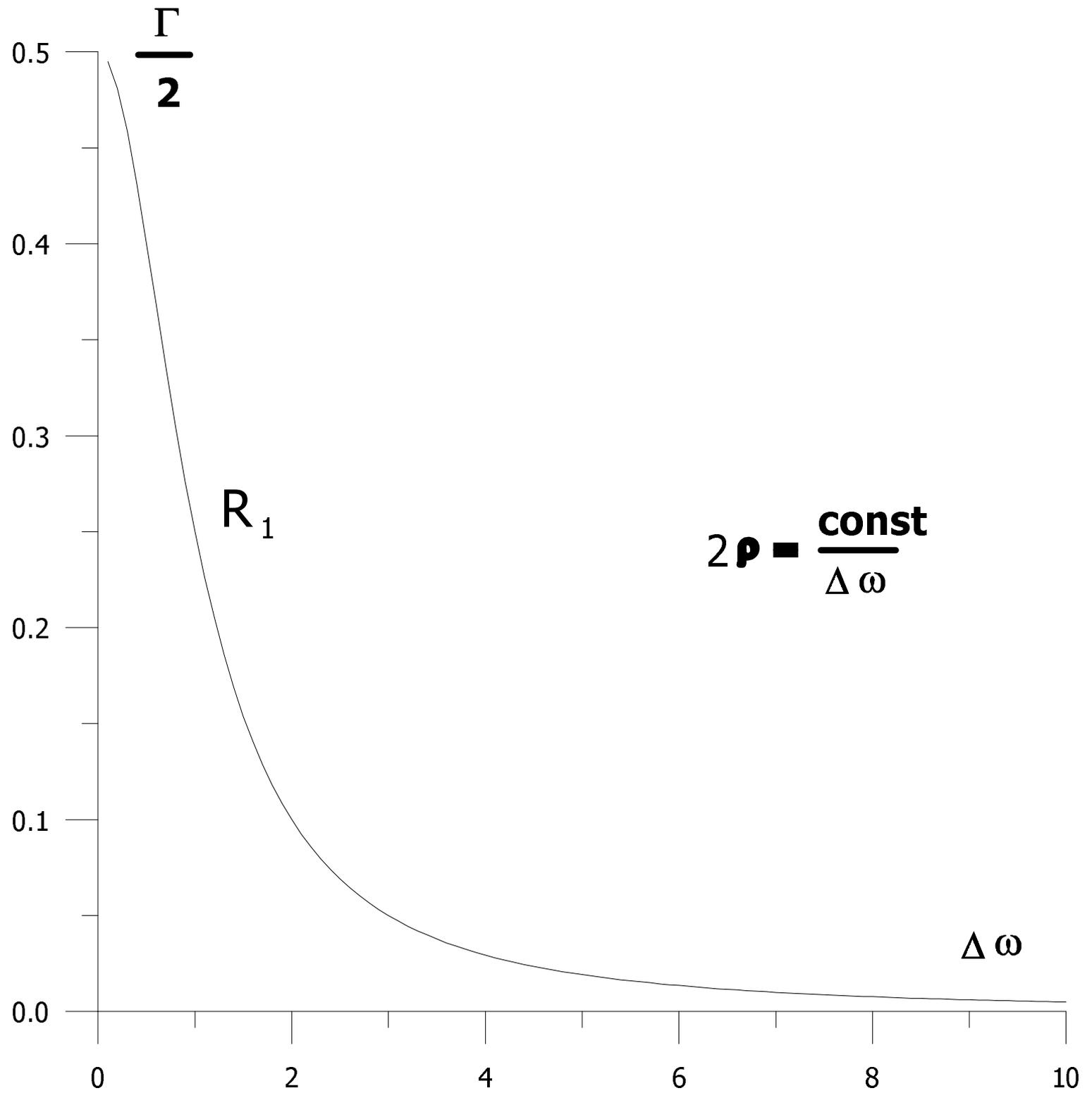
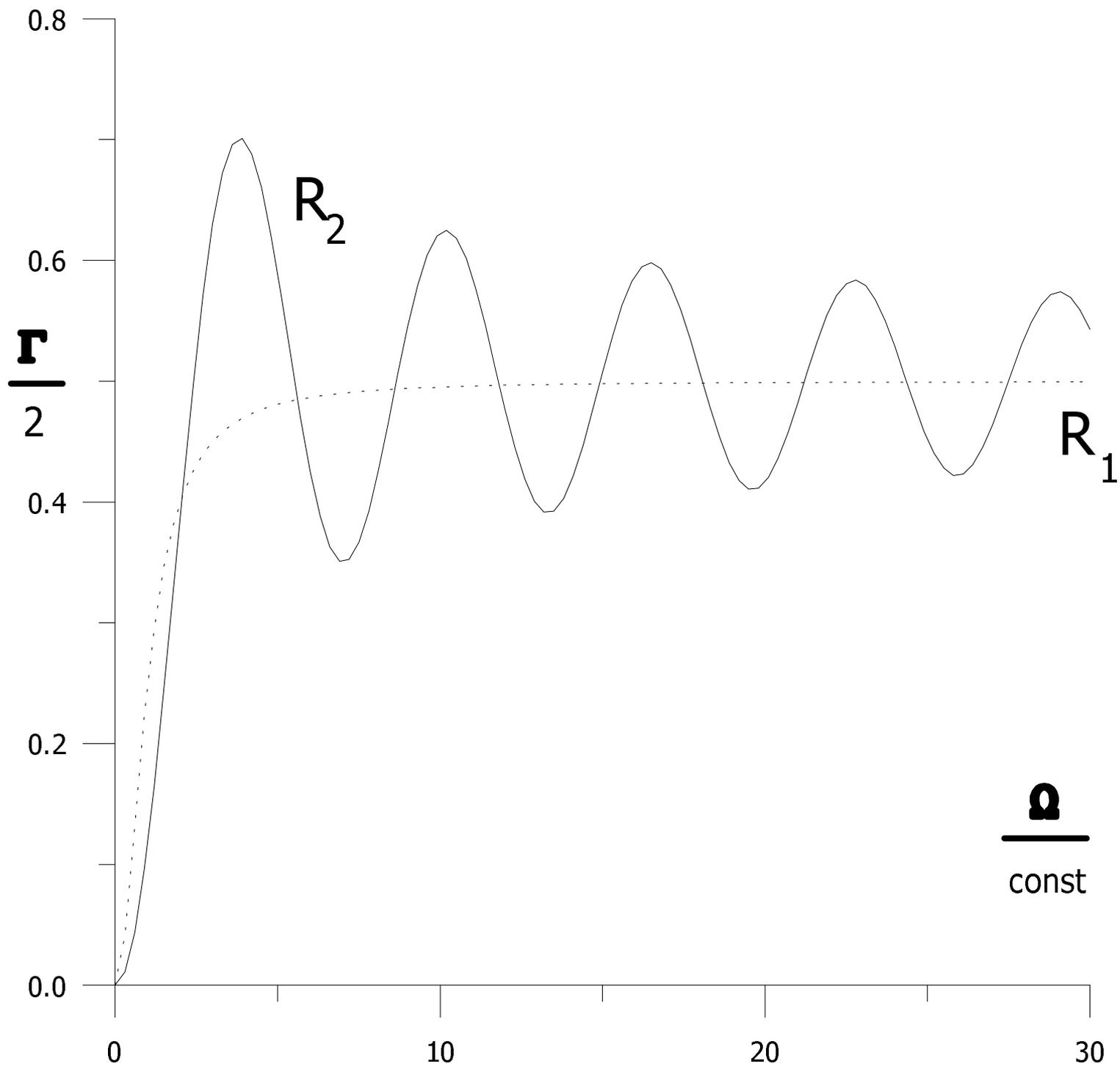
\end{document}